\documentclass[12pt,showpacs,twocolumn, preprintnumbers,aps,pre, reprint,superscriptaddress]{revtex4-1}

\usepackage{graphicx}
\usepackage{subfigure}
\usepackage{color}
\usepackage{amsmath}
\usepackage[linktocpage,colorlinks=true,linkcolor=blue,citecolor=blue,breaklinks=true]{hyperref}
\usepackage{verbatim}
\usepackage{breakcites}

\begin{document}

\preprint{}

\title{The statistical properties of sea ice velocity fields}

\author{Sahil Agarwal}
\affiliation{Program in Applied Mathematics, Yale University, New Haven, USA}
\email[]{sahil.agarwal@yale.edu}

\author{J. S. Wettlaufer}
\affiliation{Program in Applied Mathematics, Yale University, New Haven, USA}
\affiliation{Departments of Geology \& Geophysics, Mathematics and Physics, Yale University, New Haven, USA}
\affiliation{Mathematical Institute, University of Oxford, Oxford, UK}
\affiliation{Nordita, Royal Institute of Technology and Stockholm University, SE-10691 Stockholm, Sweden}
\email[]{john.wettlaufer@yale.edu}

\date{\today}

\begin{abstract}
By arguing that the surface pressure field over the Arctic Ocean can be treated as an isotropic, stationary, homogeneous, Gaussian random field, Thorndike estimated a number of covariance functions from two years of data (1979 and 1980).  Given the active interest in changes of general circulation quantities and indices in the polar regions during the recent few decades,  the spatial correlations in sea ice velocity fields are of particular interest.  It is thus natural to ask; ``how persistent are these correlations?''
To this end, a multi-fractal stochastic treatment is developed to analyze observed Arctic sea ice velocity fields from satellites and buoys for the period 1978 - 2015.  Having previously found that the Arctic Equivalent Ice Extent (EIE) has a white noise structure on annual to bi-annual time scales, the connection between EIE and ice motion is assessed. The long-term stationarity of the spatial correlation structure of the velocity fields, and the robustness of their white noise structure on multiple time scales is demonstrated, which (a) combine to explain the white noise characteristics of the EIE on annual to biannual time scales, and (b) explain why the fluctuations in the ice velocity are proportional to fluctuations in the geostrophic winds on time scales of days to months. Moreover, it is shown that the statistical structure of these two quantities is commensurate from days up to years, which may be related to the increasing prevalence of free drift in the ice pack. 
\end{abstract}

\pacs{}

\maketitle

\section{Introduction}

Polar amplification posits that if the 
 average global temperature increases, the relative change in the polar regions will be larger, and hence the observed decline of the Arctic sea ice cover during the satellite era has been a key focus of research {\cite[][]{OneWatt, Stroeve:2012}}.  
The Arctic Oscillation (AO) is an indicator of how atmospheric circulation can be related to observed changes in the sea ice cover.  However, because it captures only approximately 50\% of the variability of the sea level pressure \cite[][]{Rigor:2002}, it has been argued that the characteristics of the AO may have changed over time in a manner that the AO index is less predictive \cite[][]{Stroeve:2011}.  Of particular relevance here is the prevalence of free drift, and hence how sea level pressure and ice velocity are correlated.  First, as the ice cover has thinned, modeling studies indicate that the mechanical and dynamical properties will change, and predict that free drift will become increasingly prevalent {\cite[][]{Hakkinen:2008aa, Rampal:2009, Spreen:2011aa, Zhang:2012, Kwok:2013aa}}.  Second, the wind-driven circulation has oscillated between cyclonic and anticyclonic (at 5 to 7 year intervals) from 1948 to 1996, after which the anticyclonic pattern has prevailed \cite[][]{Proshutinsky:2015}.  Thus, a central question concerns the coexistence of the changes in circulation patterns  with the persistence of correlations between the wind and ice velocity fields.

One of a number of the feedbacks often posited to drive polar amplification is the ice-albedo feedback {\cite[e.g.,][]{Budyko:1969, Curry:1995}}.  
Due to the seasonality of the solar insolation at high latitudes, one can distill two key processes regulating the stability of the ice cover on the seasonal time scale; the ice-albedo feedback during the summer, and the loss of heat from the surface by long-wave radiation in the winter \cite[][]{MW:2011}, and these are modulated by stochastic variability \cite[][]{MW:2013}.   On multiple time scales (from weather to decadal) we have extracted this variability quantitatively from satellite data on both the ice albedo itself and the equivalent ice extent (EIE) \cite[][]{Sahil:MF},\footnote{The equivalent ice extent (EIE) is defined as the total surface area, including land, north of the zonal-mean ice edge latitude, and thus is proportional to the sine of the ice edge latitude \cite[][]{IanGeom}. Relative to the ice extent itself, the EIE minimizes coastal effects.}; two key quantities reflecting the ice-albedo feedback.   This analysis shows that the EIE and the ice albedo are multi-fractal in time rather than an AR-1 process, 
which is commonly used to characterize Arctic sea ice in climate models. Indeed, one can show that an AR-1 process is inappropriate for two key reasons; $(a)$ The existence of multiple time scales in the data cannot be treated in a quantitatively consistent manner with a single decay time for the autocorrelation, and $(b)$ the strength of the seasonal cycle is such that, if not appropriately removed, model output or satellite retrievals will always have a single characteristic time of approximately 1 year; a time scale at which all moments of the multi-fractal analysis are forced to converge \cite[][]{Sahil:MF}.  

Here, we find that the velocity field of sea ice is also multi-fractal in time \cite[][]{Rothrock:1980,Rothrock84} exhibiting points $(a)$ and $(b)$ described above.   Moreover, we find (1) a three and a half decade stationarity in the spatial correlations of the horizontal velocity components and the shear in the geostrophic wind field, yielding ostensibly the same results for 1978 - 2015 as found by \citet{Thorndike:1982aa, Thorndike:1986aa} over a two year  time window (1979 - 1980), and (2) a robust white noise structure present in the velocity fields on annual to bi-annual time scales, which we argue underlies the white noise characteristics of the EIE on these time scales.  Finally, whereas previous analyses \cite[][]{Thorndike:1982} have shown the correlation between ice motion and geostrophic wind from days to months, we find this to extend up to years.  

\section{Data and Methods}

\begin{figure}[t!]
    \centering
    \includegraphics[trim = 50 50 20 20, clip, width = 0.5\textwidth]{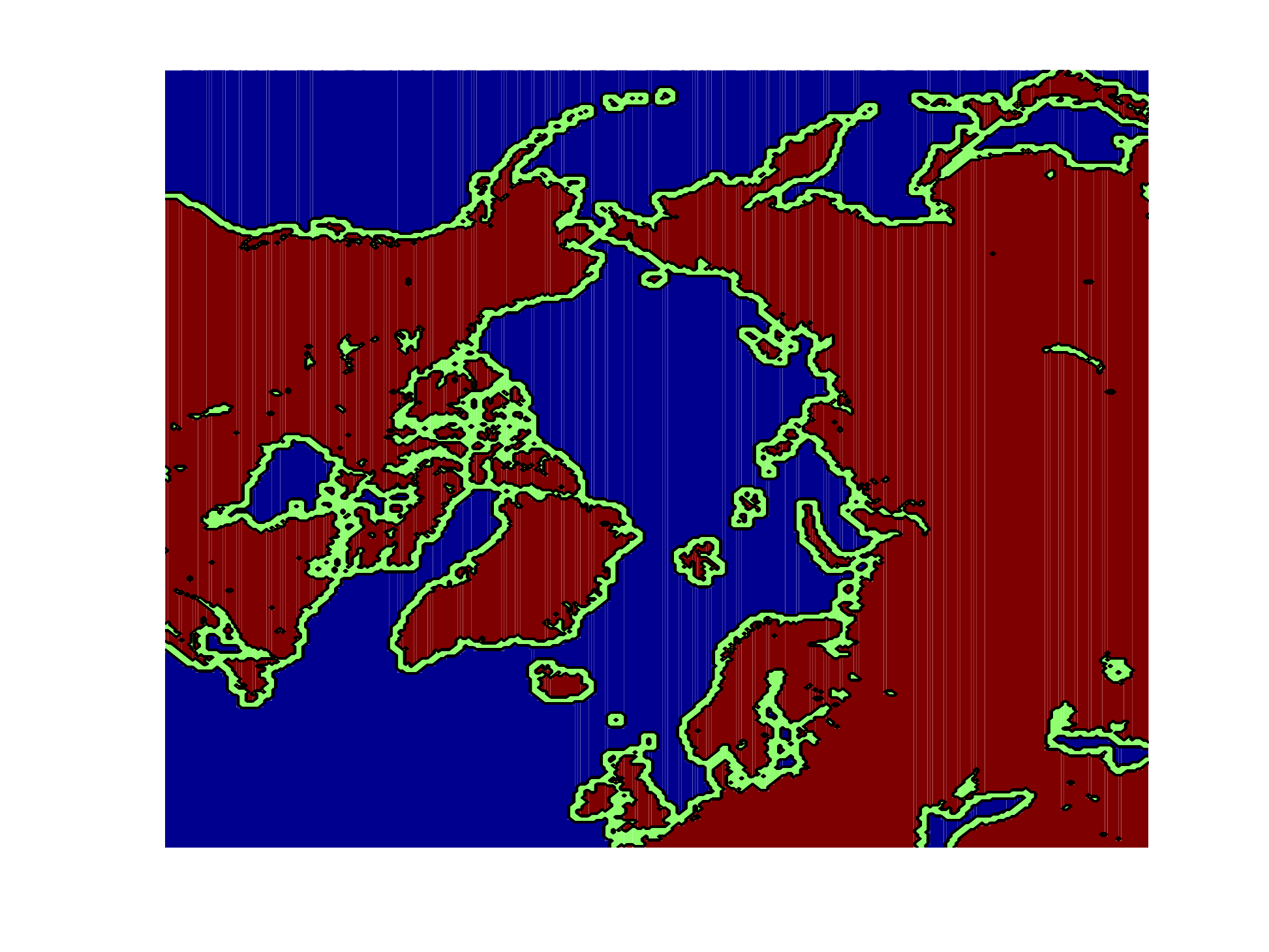}	  	
    \caption{The blue region denotes that in which the ice velocity fields are calculated, the green color denotes the region that is within 100 km of the coast, and the brown region is landmass}
    \label{fig:Arctic_map}
\end{figure}

We use the buoy derived pressure fields computed on a regular latitude-longitude grid of $2^{\circ} \times 10^{\circ}$  for the period January 1, 1979 - December 31, 2006 \cite[][]{IABP}. The ice motion velocity vectors are obtained in a gridded format from the National Snow and Ice Data Center (NSIDC) {\cite[][]{NSIDC:IceVelocity2016}}. These vectors are derived from multiple sensors that include AMSR-E, AVHRR, IABP Buoys, SMMR, SSM/I, SSMIS and NCEP/NCAR, and their coverage extends from {October 25, 1978 - May 31, 2015}. The raw ice velocity vectors from each source are processed to form the daily gridded ice velocity fields with a spatial resolution of 25 km. To minimize the effect of the coastline on ice motion, we discard all the grid points in the fields that are within a distance of 100 km of the coast \cite[e.g., see][]{Thorndike:1982, Rampal:2009} (Figure \ref{fig:Arctic_map}). 
The gridded ice motion fields have the $x$-component referenced to $90^{\circ}$ east longitude and the $y$-component referenced to $180^{\circ}$ east longitude. We calculate  the mean velocity in both the $x$- and $y$-directions for each day. We then analyze these time series using the Multi-Fractal Temporally Weighted Detrended Fluctuation Analysis (MF-TWDFA) methodology \cite[][]{Sahil:MF}, described in the next sub-section, to extract the time scales in these time series and relate them to the time scales obtained from the analysis of EIE. It is interesting to note that the ice motion data reflects both the shorter synoptic time scales and a strong seasonal cycle. Importantly, this also demonstrates that by solely looking at the bare time series, one cannot necessarily extract information regarding the process leading to such  multiple time scale fields, which emboldens us in the use of multi-fractal methods.  

\subsection{Multi-fractal Temporally Weighted Detrended Fluctuation Analysis\label{sec:DFA}}

In geophysical time series analysis, the two-point autocorrelation function is typically used to estimate the correlation time scale. This estimation has major drawbacks since ($i$) it assumes that there is only a single correlation structure in the data,  $C(t) \propto t^{-\gamma}$, where $C(t)$ is the two point autocorrelation function, ($ii$)  long term trends (linear or non-linear)  and periodicities present in the data may obscure the estimated time scales. Thus, in order to characterize the dynamics of the system, one may need multiple exponents, $\gamma$.  Namely, it is possible that for some $t \le t_{0}$ there is an exponent $\gamma_1$, for some $t_{0} < t \le t_1$ there is an exponent $\gamma_2$, and so forth. A two-point autocorrelation function will give a 1 year time scale for any signal with a sufficiently strong seasonal cycle and thereby mask any other time scales \cite[][]{Sahil:MF}. This also serves as a motivation for using the multi-fractal methodology for the sea ice velocity fields.

There are four stages in the implementation of MF-TWDFA \citep[][]{Sahil:MF}, which we summarize in turn. \\
({\bf 1.}) One constructs a non-stationary {\em profile} $Y(i)$ of the original time series $X_i$, which is the cumulative sum 
\begin{equation}
Y(i )\equiv \sum_{k=1}^{i} \left(X_k - \bar{X_k} \right), \qquad \text{where}\qquad  i = 1, ... , N.  \\
\label{eq:profile}
\end{equation}
({\bf 2.})  One divides the profile into $N_s = \text{int}(N/s)$ segments of equal length $s$ that do not overlap.  Excepting rare circumstances, the original time series is not an exact multiple of $s$ leaving excess segments of $Y(i)$.  These are dealt with by repeating the procedure from the end of the profile and returning to the beginning and hence creating $2 N_s$ segments.\\  
({\bf 3.}) In the standard $MF-DFA$ procedure an estimate is made of $Y(i)$ {\em within a fixed window} using $n$th order polynomial functions $y_{\nu}(i)$'s.  Here, however, a moving window that is smaller than $s$, but determined by the distance between points, is used to construct a point by point approximation to the profile, $\hat{y}_{\nu}(i)$.  We then compute the variance up ($\nu = 1,...,N_s$) and down ($\nu = N_s + 1,...,2 N_s$) the profile as
\begin{align}
\text{Var}(\nu, s) \equiv & \frac{1}{s}  \sum_{i=1}^{s} \{ Y([\nu - 1]s + i) - {\hat{y}}([\nu-1]s +i) \}^2 \nonumber \\
&\hspace{-10mm} \text{for $\nu = 1,...,N_s$, and} \nonumber \\
\nonumber \\
\text{Var}(\nu, s)  \equiv & \frac{1}{s}  \sum_{i=1}^{s} \{ Y(N-[\nu - N_s]s + i) - \nonumber \\
&\hspace{15mm}{\hat{y}}(N-[\nu-N_s]s +i)\}^2\nonumber \\
&\hspace{-10mm} \text{for $\nu =  N_s + 1,...,2 N_s$.}
\label{eq:varTW}
\end{align}

\begin{figure*}[t!]
    \centering
     \includegraphics[trim = 110 0 80 0, clip, width = 1\textwidth]{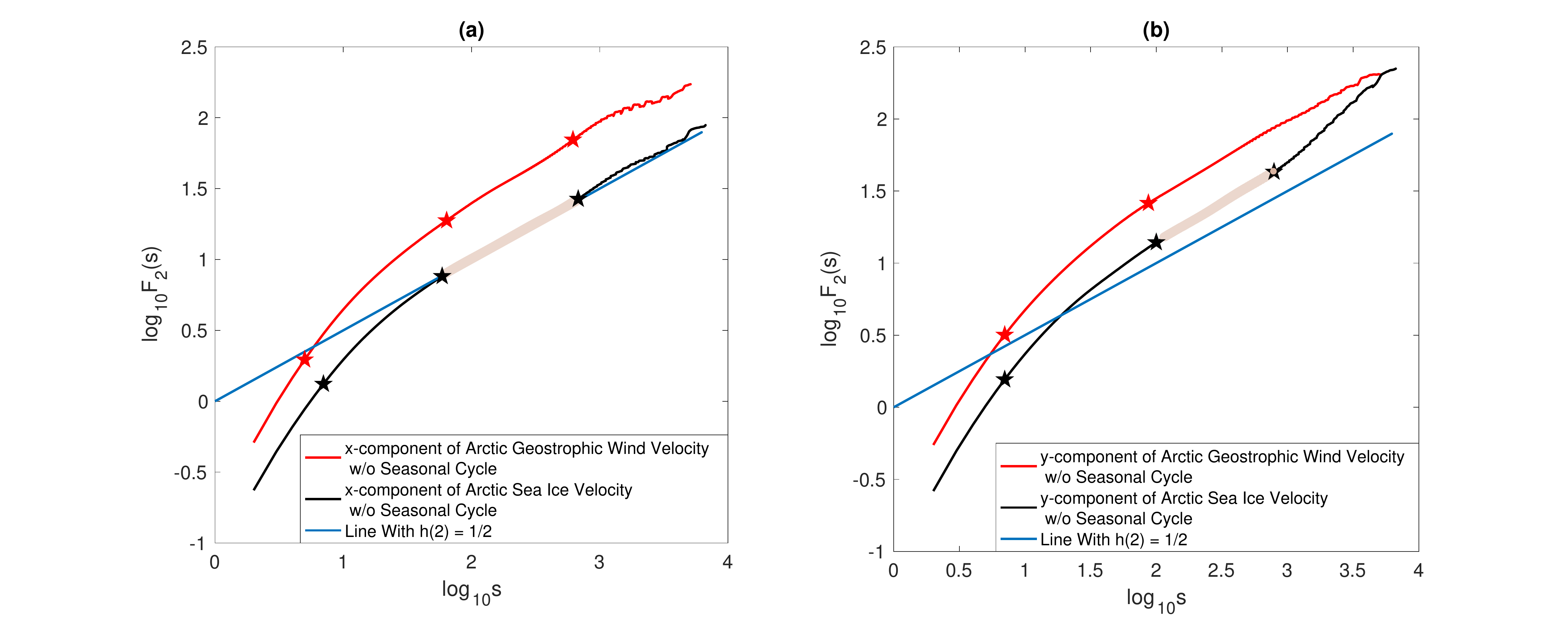}	
    \caption{The fluctuation functions $F_2(s)$ from equation \ref{eq:fluct} for daily geostrophic wind during the period 1979 - 2006 (red) and Arctic sea ice velocity during the period 1978 - 2015 (black) in the (a) \emph{x-} and (b) \emph{y-} direction after the seasonal cycle has been removed. The stars denote the crossover times associated with a change in slope (5 , 64 and 618 days for the $x-$direction geostrophic wind; 7 and 84 days for the $y-$direction geostrophic wind; 7, 60 and 680 days for the $x-$component of the sea ice velocity; and 7, 100 and 788 days for the $y-$component of the sea ice velocity). The blue line denotes white noise with $h(2)=1/2$.}
    \label{fig:Arctic_Pressure_ds}
\end{figure*}

Therefore we replace the global linear regression of fitting the polynomial $y_{\nu}(i)$ to the data, with a weighted local estimate $\hat{y}_{\nu}(i)$ determined by the proximity of points $j$ to the point $i$ in the time series such that $\vert i - j \vert \le s$.  A larger (or smaller) weight $w_{ij}$ is given to $\hat{y}_{\nu}(i)$ according to whether $\vert i - j \vert $ is small (large) \citep[][]{Sahil:MF}. \\
({\bf 4.}) The generalized fluctuation function is formed as
\begin{equation}
F_q (s) \equiv \left[ \frac{1}{2 N_s} \sum_{\nu=1}^{2 N_s} \{ \text{Var}(\nu, s)\}^{q/2} \right]^{1/q}.    
\label{eq:fluct}
\end{equation}
The principal tool of the approach is to examine how $F_q (s)$ depends on the choice of time segment $s$ for a given order $q$ of the moment taken. The scaling of  $F_q (s)$ is characterized by a generalized Hurst exponent $h(q)$ viz., 
\begin{equation}
F_q (s) \propto s^{h(q)} .  
\label{eq:power}
\end{equation}

A few characteristics are worth pointing out at this juncture \citep[e.g., ref.][and refs. therein]{Sahil:MF}. The dominant time scales are the points where the fluctuation function changes slope, i.e. shifts from one dynamical behavior to another. For a monofractal time series, the generalized Hurst exponents $h(q)$ are independent of $q$. If there is long term persistence in the data then, $h(2) = 1 - \gamma/2$ for $0 < \gamma < 1$ which also serves as a check for the two-point autocorrelation function. One can relate $h(2)$ to the slope of the power spectrum $\beta$ as follows.  If $S(f) \propto f^{- \beta}$, with frequency $f$, then $h(2) = (1 + \beta)/2$ \citep[e.g.,][]{Ding}. For a white noise process $\beta = 0$ and hence $h(2) = 1/2$. For a red noise process $\beta = 2$ and hence $h(2) = 3/2$. Therefore, the slope of the fluctuation function curves as a function of $s$ reveal the different dynamical processes that operate on different time scales. This then implies that if the data is only short term correlated ($\gamma > 1$), its asymptotic behavior will be given by $h(2) = 1/2$.  (Clearly, ``short'' and ``long'' depend in the details of the particular time series.) 
Other advantages of using temporally weighted fitting with moving windows over the regular MF-DFA are that the approximated profile is continuous across windows, reducing spurious slope changes at longer time scales, and while MF-DFA can only produce time scales up to $N/4$, MF-TWDFA extends this to $N/2$.

\begin{figure*}[t!]
    \centering
    \includegraphics[trim = 80 0 70 0, clip, width = 1\textwidth]{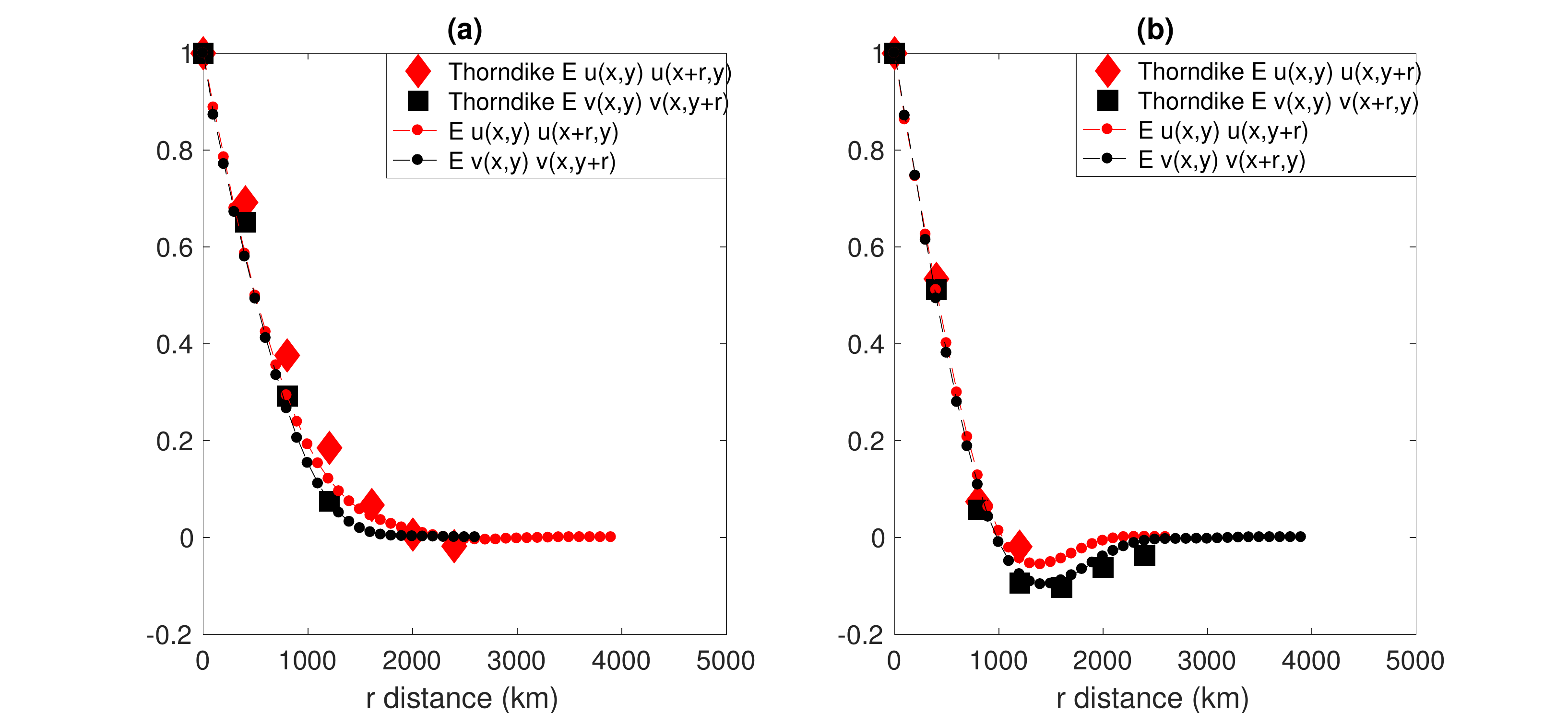}	
    \caption{The normalized spatial correlation function for Arctic sea ice velocity $u(x,y)$ and $v(x,y)$, at zero time lag, parallel (a) and perpendicular (b) to a line joining two points separated by a distance $r$, from \citet{Thorndike:1982aa, Thorndike:1986aa} and from our analysis of the data from 1978 - 2015, labeled in the legend.}
    \label{fig:Arctic_VelocityCorr}
\end{figure*}

The fidelity of regular MF-DFA was tested previously using EIE data, and it was not able to capture time scales longer than 2 years, even when 9$^{\text{th}}$ order polynomials were used to approximate the profile \citep[][]{Sahil:MF}. The principal reason for the limits on capturing time scales reside in (a) discontinuities in the profile, which lead to ``jagged'' fluctuation functions and (b) the method only provides information up to $N/4$, where $N$ is the length of the time series. Moreover, the intuition in MF-TWDFA, that points closer in time are more highly correlated than those farther apart, cannot be borne out in MF-DFA, which gives the same weight to all the points in a time window, and hence can produce spurious results for longer time scales.

\section{Results\label{sec:Disc}}

\citet{Lemke:1986} outlines the many time scales in the air/sea/ice system, ranging from days to several months, associated with the interaction of the ice with the atmosphere and the ocean mixed layer, whereas time scales ranging from decades to much longer, are ascribed to the deep ocean component of the system, generally understood to be uncoupled from {ice} drift and surface currents.
For the central basin pack ice,  the  principal forces balance air and water stresses, sea surface tilt, and the Coriolis effect \cite[][]{Thorndike:1982}. Based on 2 years of data \citet{Thorndike:1982} concluded that on time scales of days to months, more than 70\% of the variance of the ice motion can be explained by the geostrophic winds. 

In Figure \ref{fig:Arctic_Pressure_ds} we plot the fluctuation function for Arctic sea ice and geostrophic wind velocities in \emph{x-} and \emph{y-} directions without the seasonal cycle, for the period 1978 - 2015 and 1979 - 2006 respectively.
The match of the fluctuation functions for the sea ice velocity and the geostrophic wind is excellent up to characteristic (crossover) time scales of several years.  This demonstrates that the high correlation between the winds and the ice velocity concluded by \citet{Thorndike:1982} is extended from months to years, and that this correlation persists over climatological time scales.  We find that the slight difference between the fluctuation functions at longer time scales for sea-ice velocity and geostrophic winds in the $y-$direction is due to sea-ice export in the Atlantic sector via Fram Strait. We determined this by comparing the fluctuation function (not shown here) for the pressure difference between the Arctic basin and the $50-60^\circ{N}$, $20^\circ{W} - 20^\circ{E}$ region, with the sea-ice velocity in the $y-$direction. These time scales are the same as those of the North Atlantic Oscillation.

Previously we showed that when the seasonal cycle is removed from the EIE over the 32 year period (1978-2010), it exhibits a white noise dynamical behavior on annual to bi-annual time scales \cite[][]{Sahil:MF}. Here, we extend the analysis of the EIE without the seasonal cycle and examine it in progressive periods; 1978-1980, 1978-1981, and 1978-2014. This analysis confirms the presence of white noise structure as a robust signal on annual to bi-annual time scales. Finally, the data from a hybridized data set from 1901-2014 compiled by Walsh and Chapman \cite[][]{Walsh_Chapman} also shows white noise structure on these time scales. Thus, for such a robust signal to exist, the physical mechanism responsible for it must be stationary. 

\begin{figure}[htbp]
    \centering
        \includegraphics[trim = 40 0 20 0, clip, width = 0.5\textwidth]{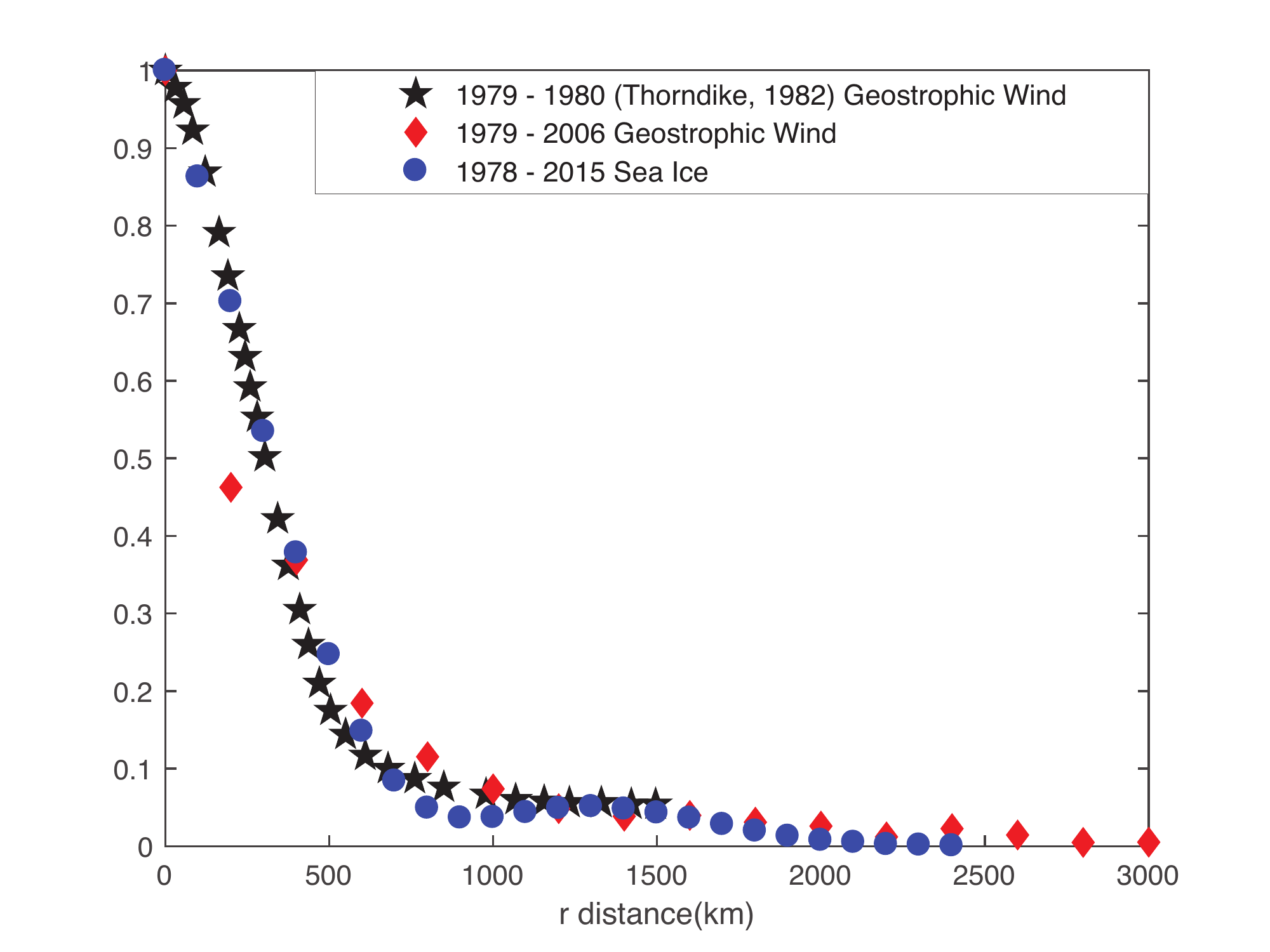}	  	
    \caption{The normalized spatial correlation function for shear in the {geostrophic wind field \cite[from][]{Thorndike:1982aa}} for 1979 - 1980 (pentagram, black) and for 1979 - 2006 (diamond, red). To see the structural relationship between the wind field and the sea ice motion field, we show the normalized spatial correlation function for shear in the Arctic sea ice velocity field for 1978 - 2015 (circle, blue).}
    \label{fig:Ice_Shear}
\end{figure}

\begin{figure}[htbp]
    \centering
        \includegraphics[trim = 40 0 20 0, clip, width = 0.5\textwidth]{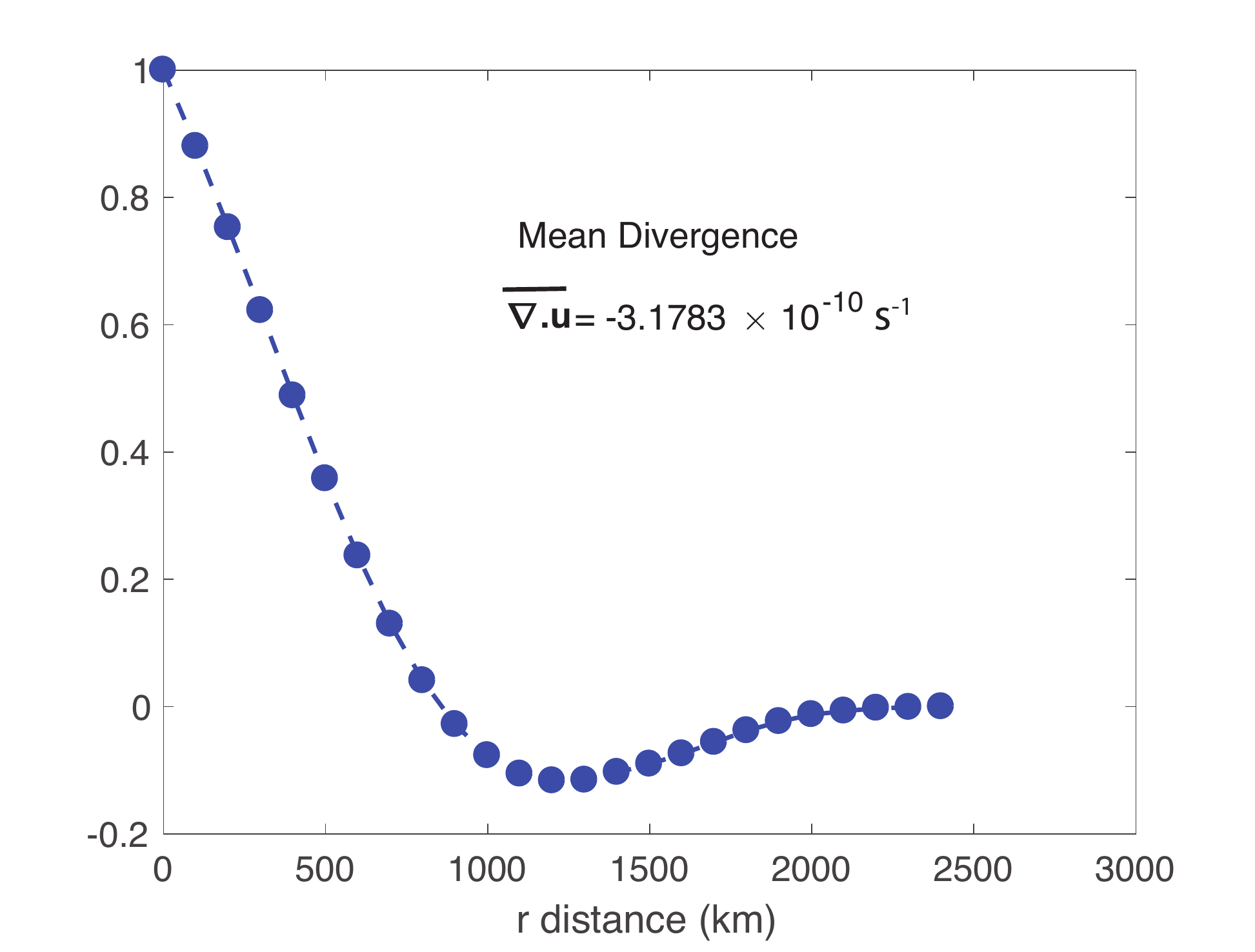}	  	
    \caption{The normalized spatial correlation function for divergence of the sea ice velocity fields, for the standard case treated, which avoids regions within 100 km of the coast. The temporal mean of the divergence field is $\overline{\nabla\cdot\bf{u}} = -3.18 \times 10^{-10}~s^{-1}$ showing that the velocity field of sea ice is nearly divergence free.  We note here (see text) that if we avoid regions within 400 km of the coast, the mean divergence decreases further, giving  $\overline{\nabla\cdot\bf{u}} = 6.3976 \times10^{-11}~s^{-1}$. }
    \label{fig:IceDiv}
\end{figure}

To further pursue the robustness of the statistics, we calculate the correlation between the components of sea ice velocity, $u(x,y), v(x,y)$ parallel or perpendicular to a line joining two points separated by a distance $r$ \cite[see e.g.,][]{Thorndike:1986aa}.  The spatial autocorrelation in both the parallel and perpendicular directions for 2 years of data (1979 - 80) \cite[from][]{Thorndike:1986aa} and for 37 years of data (1978 - 2015) are shown in Figure \ref{fig:Arctic_VelocityCorr}.  {In Figure \ref{fig:Ice_Shear} we compare the spatial autocorrelation functions in the shear for the (i) geostrophic winds from 1979 - 1980 \cite[from][]{Thorndike:1982aa}, (ii) ice velocity from 1978 - 2015, and (iii) geostrophic winds from 1979 - 2006}.  These demonstrate a striking three decade stationarity in the key correlations underlying the structure of the velocity field of pack ice. { Figure \ref{fig:IceDiv} shows the spatial structure of the divergence field for the complete record, with the inset showing the temporal mean of the sea-ice velocity divergence field. The sea-ice velocity field exhibits solenoidal behavior over the entire domain, with the spatial correlation reaching a minimum at $\approx$1200 km.  To study the effect of the coast on the dynamics of the sea-ice velocity fields, we compared the fluctuation functions and the spatial correlation functions when regions within 400 km of the coastline were discarded.
We find that the while the magnitude of the fluctuation functions change, the slopes do not and these encode all of the dynamical information.  
The increase in magnitude of the fluctuations is attributed to an increase in the magnitude of sea-ice velocities with distance from the coast. This attribution is supported by the facts that (1) the spatial correlation functions are the same for the 400 km and the 100 km threshold, and (2) the mean divergence decreases as we discard more ice area at the basin boundary, so that $\overline{\nabla\cdot\bf{u}} = 6.3976 \times10^{-11}~s^{-1}$ for regions at least 400 km away from the coast \cite[for comparison, see Table 5 in][]{Thorndike:1986aa}}

\begin{figure}[h!]
    \centering
    \includegraphics[trim = 20 0 20 0, clip, width = 0.5\textwidth]{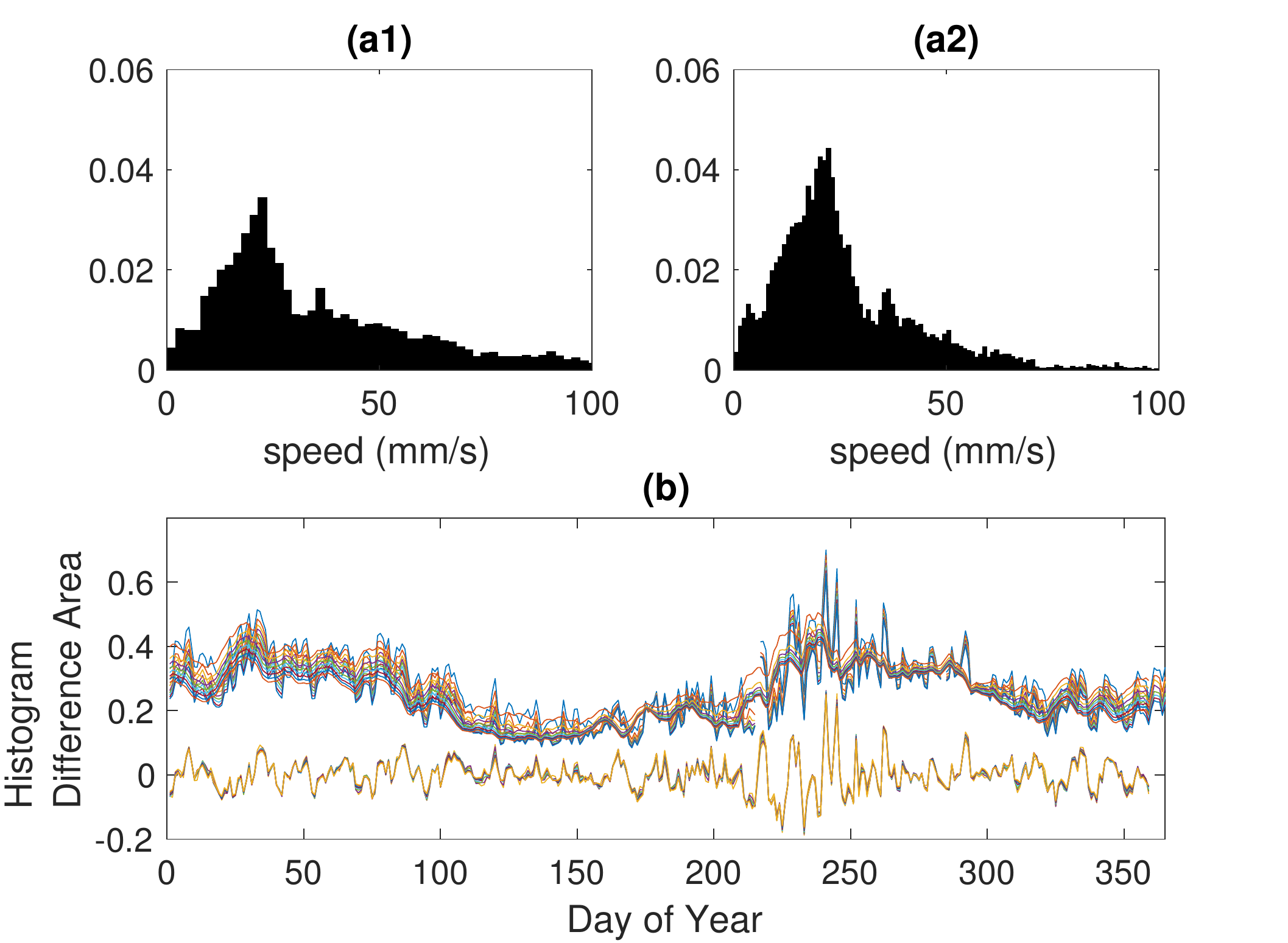}
    \caption{(a) The normalized Arctic sea ice speed histograms shown for day 1 of the year (January 1). If there is ice in a pixel for at least time $\tau$ ($\tau_\text{left} = 1 \text{yr}, \tau_\text{right} = 37 \text{yr}$), then we compute the average speed for that pixel. (b) The difference in the histograms for each day of the year for $\tau = 1, 2, ... 12 \text{yr}$ with respect to $\tau = 37\text{yr}$. The bottom curves denote the residual after a running mean with window of length 7 days has been removed from the top curves for different thresholds.  Clearly all of the residuals collapse to a single curve. }
    \label{fig:Arctic_VelocityThreshold}
\end{figure}

Finally, we use the entire 37 year record to calculate the mean speed of each pixel every day of the year. This mean is computed using a specified threshold, $\tau$, i.e., if a pixel has contained sea ice for $\tau$ years, where the maximum $\tau$ is 37 years.   Thus, thresholds specify a minimum time for which a pixel contained ice \cite[][]{Sahil:grl}. These are then used to produce a histogram for each day of the year for the sea ice speed, with each bin representing the number of pixels having the corresponding speed. These histograms are then normalized in order to compare between different thresholds. Figure \ref{fig:Arctic_VelocityThreshold}(a) shows the normalized histograms for January 1. The left histogram is for $\tau = 1$ year, {Fig. \ref{fig:Arctic_VelocityThreshold}(a1)}, and the right histogram is for $\tau = 37\text{ years}$, {Fig. \ref{fig:Arctic_VelocityThreshold}(a2)}. Next, for each day of the year we compute the difference in the two histograms, calculate the area under the curve, and plot this in Figure \ref{fig:Arctic_VelocityThreshold}(b) for different thresholds from $\tau = 1, 2, ..., 12~\text{years}$, with respect to $\tau = 37\text{ years}$. 

The change in the area as the threshold $\tau$ is increased is negligible, with the maximum variability appearing during the summer, as expected due to the typical seasonality of free drift. Moreover, when we subtract the running mean with a window size of 7 days from all the curves, and plot the residuals, the curves for all thresholds collapse (lower curve in Fig. \ref{fig:Arctic_VelocityThreshold}b).  Finally, when we analyze the residuals with the MF-TWDFA methodology, we extract the weather time scale of 10 days and an approximately 47 day time scale, associated with the high variability during the summer as seen in Fig. \ref{fig:Arctic_VelocityThreshold}(b). These results further demonstrate the stationarity of the Arctic sea ice velocity fields.  We therefore ascribe the white noise structure of the EIE to that of the velocity field.

\section{Stochastic Model for Sea Ice Velocity \label{sec:vel}}

The multiple time scales of Arctic sea ice motion vectors extracted from the {multifractal} analysis have been described in \S\ref{sec:Disc}.  In all cases, we can attribute the $\sim$ 5 day time scale to the relaxation time scale for synoptic fluctuations in the sea ice motion \cite[][]{Thorndike:1986ab}.  When the seasonal cycle is not removed, the only time scales are due to synoptic scale weather and the seasonal cycle itself. When the seasonal cycle is removed, the slope of the fluctuation curves demonstrate that a white noise process operates on time scales $\gtrsim$ 60 days.  The data motivate a stochastic treatment of the ice motion, which for simplicity of illustration we write for the $u$-component of the velocity vector 
as
\begin{equation}
\frac{d}{dt}u(t) = -\frac{1}{\cal T}u(t) + \sigma_{1}\xi(t) + \sigma_{2}\sin(\omega t), 
\label{eq:Lngvn}
\end{equation}
where $u(t)$ is the daily average ice velocity component, ${\cal T}$ is the relaxation time scale, $\omega = {2 \pi}/{365}$ is the frequency of the seasonal cycle, and $\sigma_{1}$ and $\sigma_{2}$ are the strengths of the respective forcings.

\begin{figure}
        \centering
        \includegraphics[trim = 15 0 20 0, clip, width = 0.5\textwidth]{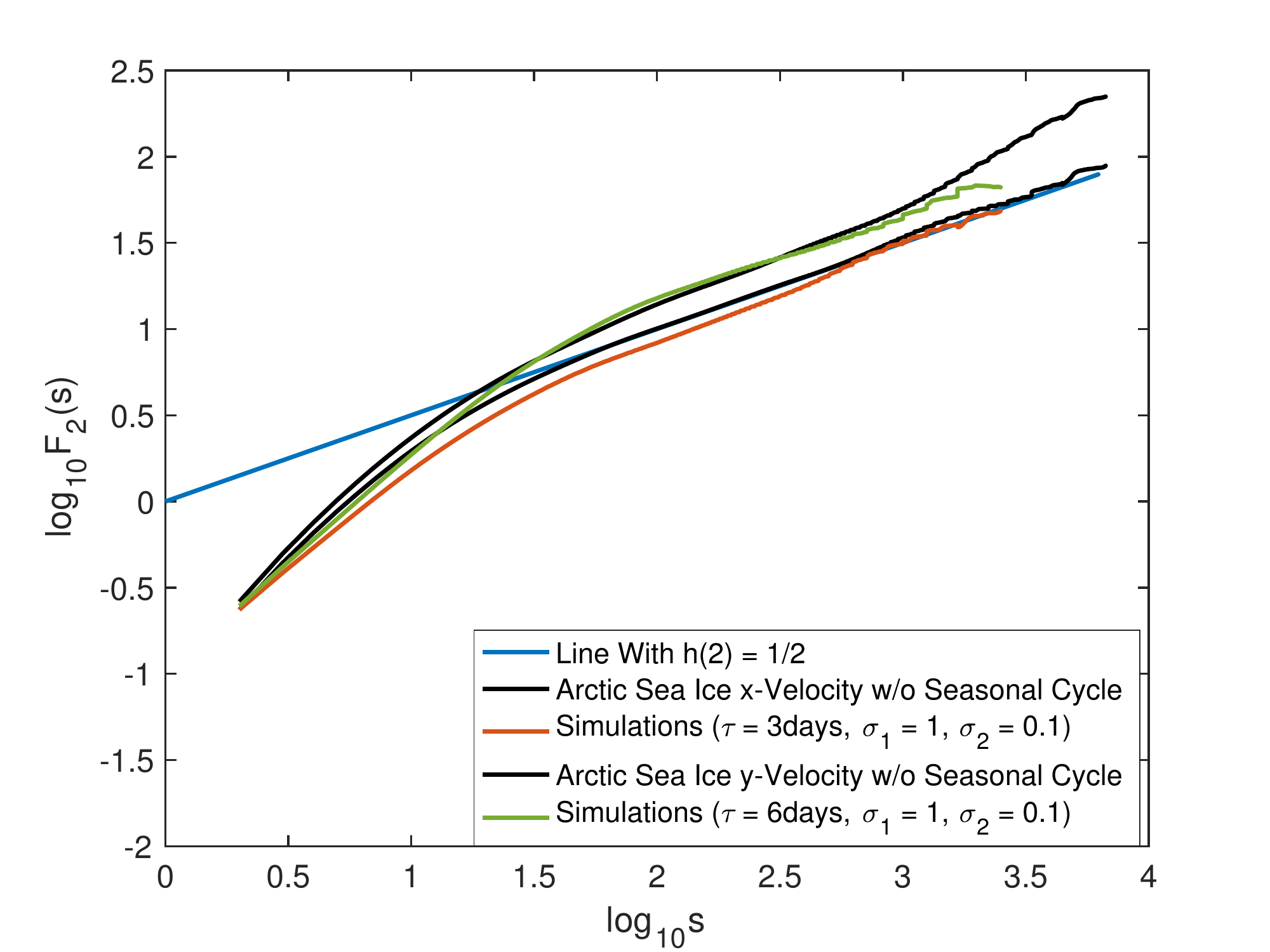}	
    \caption{The model described by Eq. (\ref{eq:Lngvn}), with $\sigma_{1} = 1$ and $\sigma_{2} = 0.1$, is compared with observations of the $u$ and $v$ components of Arctic sea ice velocity without the seasonal cycle.}
    \label{fig:LngvnCompare}
\end{figure}

Figure \ref{fig:LngvnCompare} shows the model to be in excellent agreement with the observations for time periods of days to decades.   From periods of months to decades the dynamics is ostensibly white and thus 
the variance of the ice velocity is quasi-stationary.   This is the key point, as the comparison is robust for reasonable changes in the model parameters (e.g., $\cal T$ ranging from 2 to 6 days).  However, the model is a minimal one and by considering a variety of other effects, such as multiplicative noise  in combination with the periodic forcing, the potential for a stochastic resonance arises \cite[][]{MW:2013}, which might produce behavior not found in the observational record.  However, it is the observations themselves, with their long term stationarity, that motivate the simplicity of the model and thus its utility.  
%

\section{Conclusion}

Using a variety of  stochastic analysis approaches we examined approximately three decades of data and demonstrate the stationary structure of the correlation between sea ice motion and geostrophic winds.   With two years of data, \citet[][]{Thorndike:1982} showed that on time scales of days to months, more than 70\% of the variance of the ice motion { in the central basin} can be explained solely by the geostrophic winds.  Over climatological time scales, we find a striking robustness of this conclusion, which extends from days to years, and thus is most likely associated with the prevalence of free drift as the ice cover has declined \cite[e.g.,][]{Zhang:2012}. 
We find that the ice motion field exhibits a white noise structure that explains that found in the Equivalent Ice Extent over the same three and a half decade period.  This is due to the long-term stationarity of the spatial correlation structure of the velocity fields.  {Moreover, the sea-ice velocity field exhibits solenoidal behavior over the entire domain.} Finally, using a periodically forced stochastic model, we can explain the combination of time scales that underlie the observed persistent structure of the velocity field and the forcing that produces it.  These results can act as a test bed for the statistical structure of model results.  



\acknowledgments

The authors acknowledge NASA Grant NNH13ZDA001N-CRYO for support.  JSW acknowledges Swedish Research Council grant no. 638-2013-9243 and a Royal Society Wolfson Research Merit Award for support.

\section*{Appendix A}
\subsection*{Optimal Interpolation}


To obtain the spatial correlation functions, optimal interpolation is performed to interpolate the data on a rectangular grid using the following algorithm \citep{Daley:1993aa}:

\begin{itemize}
\item Calculate the background field $\mathbf{{\bar{F}}_b}$ by first performing cubic interpolation on the raw data to analyze grid points for all days.  Then take the mean field. \\
Note: Using a constant or a non-climatological field would be unsuitable as the results would be poorly constrained \citep{Thorndike:1982}.
\item Calculate the background error correlation matrix $\mathbf{\bar{\bar{B}}}$ as 
\begin{equation}
B(i,j) = \text{exp} \left[-\left(\frac{\Delta r}{L} \right)^2 \right],
\label{eq:bg}
\end{equation}
where $\Delta r = \sqrt{[x(i) - x(j)]^2 + [y(i) - y(j)]^2}$. For the pressure field $L = 250$ km, for the ice velocity field $L = 2000$ km.

\item Calculate the weights $\mathbf{\bar{w}}$ as 
\begin{equation}
\left(\mathbf{\bar{\bar{B}}} + \frac{\sigma ^2}{\eta ^2}\right) \mathbf{\bar{w}} = \mathbf{\bar{B}_{xy}},
\label{eq:wt}
\end{equation}
where $\sigma^2 = 1 \text{mb}^2$ is the variance of the observations, $\eta^2 = 9 \text{mb}^2$ is the variance of the background field, and $\mathbf{\bar{B}_{xy}}$ is the background error correlation vector for analysis grid point $(x,y)$ to the location of the observations.

\item Interpolate the background field at the location of the observations to obtain $\mathbf{{\bar{F}}_{bo}}$.
\item The interpolated field is then obtained as
\begin{equation}
\mathbf{F_a(x,y)} = \mathbf{F_b(x,y)} + \mathbf{\bar{w}^T}(\mathbf{\bar{F}_o} - \mathbf{{\bar{F}}_{bo}}).
\label{eq:oi}
\end{equation}
The above algorithm assumes that the observation and background errors are uncorrelated with zero mean.

\item The spatial derivatives for pressure are then calculated using forward differences at each grid point \citep{IABPreport}:
\begin{equation}
\frac{\partial p}{\partial x} = \frac{p(x+h) - p(x)}{h},
\label{eq:pd}
\end{equation}
where $h = 1$km. Similarly, other derivatives $\frac{\partial p}{\partial y}, \frac{\partial^2 p}{\partial x^2}$, etc., are calculated.
\item Finally, in order to calculate the spatial autocorrelation function for field $\mathbf{\bar{\bar{T}}}$, the mean quantity for the whole record is removed, i.e. the function is calculated for $\mathbf{\bar{\bar{T}}}^\prime$, where
\begin{equation}
\mathbf{\bar{\bar{T}}}^\prime= \mathbf{\bar{\bar{T}}} - \widetilde{\mathbf{ \langle {\bar{\bar{T}}} \rangle}}, 
\label{eq:anomaly}
\end{equation}
and $\widetilde{\mathbf{ \left \langle {.} \right \rangle}}$ is the time average of the grid point average of the field.

\end{itemize}

\section*{Appendix B}
\subsection*{Shear and Divergence in Velocity Fields and Spatial Correlation Functions}
The horizontal shear in the geostrophic winds, as well as the sea-ice velocity fields is calculated as 
\begin{equation}
\xi = \left [ \left ( \frac{\partial v}{\partial x} + \frac{\partial u}{\partial y} \right )^2 + \left ( \frac{\partial u}{\partial x} - \frac{\partial v}{\partial y} \right )^2 \right ]^{1/2}, 
\label{eq:Shear}
\end{equation}
where $u$ is velocity in $x-$direction and $v$ is velocity in $y-$direction.

The divergence in the sea-ice velocity field is calculated as
\begin{equation}
\zeta = \frac{\partial u}{\partial x} + \frac{\partial v}{\partial y},
\label{eq:Div}
\end{equation}
where $u$ is velocity in $x-$direction and $v$ is velocity in $y-$direction.

The spatial autocorrelation function $\psi(r)$ for a field $\mathbf{\bar{\bar{\zeta}}}$ is calculated as follows:
\begin{itemize}
\item The normalized autocorrelation function is computed separately for each row ($x-$direction) / column ($y-$direction) of the matrix for all times.
\item A weighted average of all the autocorrelation functions computed in the previous step is then calculated with the weights given by the distance over which each respective autocorrelation function is computed; 
\begin{equation}
\psi(r) = \frac{1}{\sum_{k}L_k}\sum_{k}L_kC_k(r),
\label{eq:xcorr}
\end{equation}
where $C_k(r)$ is the normalized autocorrelation function for a row/column $k$ calculated over a distance $L_k$.
\end{itemize}


%



\end{document}